\documentclass{blois}

\def\be{\begin{equation}}
\def\ee{\end{equation}}
\def\bea{\begin{eqnarray}}
\def\eea{\end{eqnarray}}

\newcommand{\Photo}{}

\usepackage{graphicx}
\usepackage{epsfig}
\usepackage{amssymb, amsmath, mathbbol}

\begin{document}
\vspace*{4cm}
\title{Lepton flavour violation and neutrino physics: beyond the
  Standard Model \footnote{PCCF RI 16-05} }

\author{Ana M. Teixeira}

\address{Laboratoire de Physique Corpusculaire, CNRS/IN2P3 -- UMR 6533,\\
Campus Universitaire des C\'ezeaux, 4 Av. Blaise Pascal,
63178 Aubi\`ere Cedex, France}

\maketitle\abstracts{
If observed, charged lepton flavour violation  is a clear sign
of new physics - beyond the Standard Model minimally extended to accommodate
neutrino oscillation data. 
After a brief review of several charged lepton flavour violation
observables and their current
experimental status, we consider distinct 
extensions of the Standard Model which could potentially 
give rise to observable signals, focusing on the case of models 
in which the mechanism of neutrino mass generation is the common
source of neutral and charged lepton flavour violation.}

\section{Introduction}
Three major observations cannot find an explanation in the 
Standard Model (SM); these include the baryon asymmetry of the
Universe (BAU), the lack of a viable dark matter (DM) candidate and
finally, neutrino oscillations.

Leptonic mixings and massive neutrinos offer a true gateway
to many new experimental signals or deviations from SM predictions in
the lepton sector; among others, these include charged lepton
flavour violation (cLFV). 
The most minimal extension of the SM allowing to accommodate $\nu$
oscillation data consists in the addition of right-handed neutrinos
($\nu_R$) while preserving the total lepton number, 
thus giving rise to Dirac masses for the neutral leptons. In such a
framework, individual lepton numbers are
violated  (as encoded in the lepton mixing matrix, $U_{\rm{PMNS}}$), and 
cLFV transitions such as $\mu \to e \gamma$ can occur, being
mediated by $W^\pm$ bosons and 
massive neutrinos. However, and due to the tiny values of light neutrino
masses, the associated rate is extremely small,
BR($\mu \to e \gamma $)$ \sim \mathcal{O}(10^{-54})$, lying beyond the
reach of any future experiment. Thus, the observation of a cLFV process
would necessarily imply the existence of new physics degrees of
freedom (beyond minimal extensions via massive Dirac neutrinos). 

At present, many cLFV observables are being searched for in numerous
facilities; after a brief summary of the current status of the
experimental searches, we consider the impact of SM extensions
regarding cLFV. An interesting class of models is that in
which cLFV arises from the mechanism of $\nu$-mass generation: 
among the several models successfully accounting 
for and explaining $\nu$-data, many offer the possibility to further
address the BAU via leptogenesis, and/or succeed in putting 
forward viable DM candidates, or
even ease certain theoretical puzzles of the SM. In particular, we 
will summarise 
the prospects for cLFV observables of several appealing seesaw
realisations. 

\section{cLFV observables and facilities: brief overview}
Other than $\nu$-oscillations (by themselves signalling flavour
violation in the neutral lepton sector), lepton flavour violation  
can be searched for in a
number of rare decays and transitions, both at high-intensities and at
high-energies \footnote{For dedicated reviews,
  see~\cite{cLFV.revs}.}. Current searches have already put stringent
bounds on 
these rare processes, and the next generation of experiments is
expected to further improve them~\cite{cLFV.exp}. 

A number of cLFV observables emerges in association with the so-called
{\it muonic channels}. Radiative cLFV muon decays, $\mu^+ \to e^+ \gamma$, 
have been searched for since the 1940s; the event signature consists
of back-to-back coincident positron-photon pairs, with a well-defined
energy ($E_e = E_\gamma = m_\mu/2$). There are prompt and accidental
backgrounds to the process: while the former are related with
SM-allowed radiative muon decays ($\mu \to e \nu_e \nu_\mu \gamma$),
and scale proportional to the rate of the stopped muons, $R_\mu$, the latter
include coincidences of photons (from SM-allowed radiative decays, or
in flight $e^+ e^-$ annihilation) with a positron from Michel decays,
and typically scale with  $R_\mu^2$. The current bound on these decays 
is BR($\mu \to e \gamma$)$<4.2 \times 10^{-13}$ (MEG, 2016). In the
future, MEG II should be able to bring down the sensitivity to 
$6 \times 10^{-14}$. 

\noindent
Three-body cLFV muon decays, $\mu^+ \to e^+ e^- e^+$, are also a
previleged channel to look for new physics. The event signature is associated
with a final state composed of three charged particles, coincident and
arising from a common vertex. 
Likewise, there are
physics and accidental backgrounds, the latter being the dominant
ones. The current bound BR($\mu \to e e e $)$<1.0 \times 10^{-12}$
(SINDRUM, 1988) is expected to be ameliorated in the near future by
the Mu3e experiment ($10^{-15}$, possibly $10^{-16}$ if a very
high-intensity muon beam is available).  

{\it Muonic atoms} also offer a rich laboratory to study cLFV - these are
$1s$ bound states which are formed when a $\mu^-$ is stopped in a
target. The muon can then decay via SM-allowed processes (decay in
orbit, nuclear capture, ...) or, in the presence of new physics,
undergo cLFV transitions. An example is that of the neutrinoless $\mu-e$
conversion, $\mu^- + (A,Z) \to e^- + (A,Z) $: the rate of the
(coherent) process typically increases with the atomic number, being
maximal for $30 \leq Z \leq 60$. The event signature consists of a
single mono-energetic electron, whose energy (albeit target-dependent)
lies close to 100~MeV, being thus easilly distinguishable from the
energy of electrons arising from the Michel spectrum of free $\mu$
decays. This is a clean
process, which only has physics backgrounds (e.g. muon decay in orbit)
suffering to a minor extent from beam purity issues, cosmic rays, ... 
The best limit has been obtained for Gold nuclei, 
CR($\mu-e$, Au)$< 7 \times 10^{-13}$; in the future, several
collaborations plan to significantly improve the sensitivity
to $\mu-e$ conversion, and
these include DeeMe - CR($\mu-e$, SiC)$<10^{-14}$, Mu2e - CR($\mu-e$,
Al)$< 10^{-17}$, COMET Phase I (II) - CR($\mu-e$, Al)$< 10^{-15
  (-17)}$, and ultimately PRISM/PRIME - CR($\mu-e$, Ti)$< 10^{-18}$.

\noindent
Another interesting observable consists in the cLFV decay of a muonic
atom into a pair of electrons~\cite{Koike:2010xr}, 
$\mu^- e^- \to e^- e^-$; the 
Coulomb interaction between the muon and the electron wave functions leads
to an enhancement of the associated decay rate, which can scale
proportionally to 
$(Z-1)^3$ - or even stronger, for large $Z$ atoms (suggesting
that experimental setups with Lead or Uranium atoms could be
considered). This is a ``new'' observable, which could be included in
the Physics Programme of COMET, and also be studied at Mu2e.

\begin{figure}   
\begin{center}
\epsfig{file=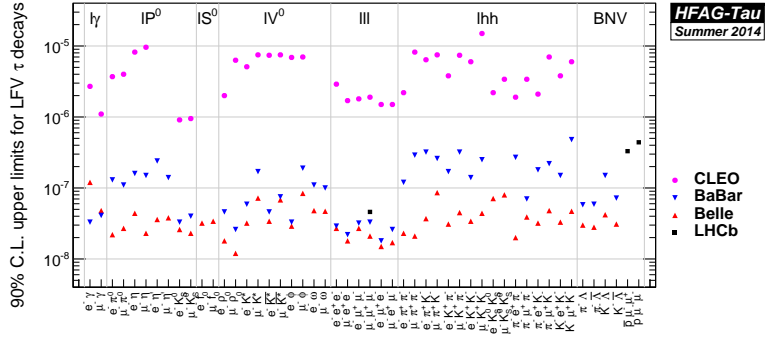, width=100mm}
\end{center}
\caption[]{Summary of the limits on several cLFV tau decays (90\% C.L.), 
as given by the HFAG-Tau 2014 Summer 
Report~\cite{Amhis:2014hma}.}\label{fig:taudecays}
\end{figure}

Due to its larger mass, 
{\it decays of the tau lepton} offer numerous channels in which to
search for cLFV. Taus can be pair-produced in $e^+ e^-$ collisions,
and events are divided into two hemispheres: one is devoted to tagging
tau leptons relying on SM decays (such as $\tau \to \nu_\tau \nu_e
e$), while the other is used to search for rare cLFV decays. Purely
leptonic cLFV processes include radiative decays 
($\tau \to \ell \gamma$) and three body final states ($\tau \to \ell_i
\ell_j \ell_k$). Event signatures are established on criteria for the
invariant mass and the total energy of the final state
(e.g. $E_{3 \ell} -\sqrt{s} /2 \sim 0$, $M_{3\ell} \sim m_\tau$).
Contrary to radiative decays, which are plagued by both physical and
accidental backgrounds, the three-body $\tau$ decays 
do not suffer from irreducible 
backgrounds. At present, the bounds for the different channels lie
around $10^{-8}$; future prospects, which
include a SuperB or a Tau-Charm factory, are expected to improve the
current sensitivities to $10^{-9}$ ($10^{-10}$) for radiative (3-body)
decays. Semi-leptonic tau decays offer further possibilities to look
for cLFV (as well as lepton number violation or even baryon number
violation); a summary of the upper limits on many of the latter
channels can be found in Fig.~\ref{fig:taudecays}. 

Abundant data on {\it leptonic and semi-leptonic meson decays} has further
allowed to constrain many cLFV decay modes. As an example, we quote
here some of the current limits~\cite{Agashe:2014kda}: 
BR($K_L \to \mu e $)$< 4.7 \times 10^{-12}$, 
BR($K^+ \to \pi^+ \mu^+ e^-$)$2.1< \times 10^{-11}$,
BR($D^0 \to \mu e $)$< 1.5\times 10^{-8}$ and
BR($B \to \mu e $)$< 2.8\times 10^{-9}$.

Finally, cLFV can also be manifest in the decays of SM bosons, such as 
$Z\to \ell_i \ell_j$, for which recent LHC bounds have already started
superseeding previous LEP bounds~\cite{Agashe:2014kda}; 
likewise, the impressive
amount of Higgs states produced in both LHC runs has also allowed to
study and constrain very rare cLFV decay modes such as $H \to \mu
\tau$. 
Should new physics states be produced at the LHC, one can naturally
look for their cLFV decays, 
possibly induced by new flavour-violating interactions
(although the properties of the final states, and the experimental
signatures, are strongly model-dependent). 

\section{cLFV and New Physics models}
Interpreting experimental data on cLFV observables 
- be it in the form of a possible
measurement or improved bounds - requires an underlying theoretical
framework: new physics models can lead to ``observable'' cLFV
introducing new sources of lepton flavour violation, as well as new operators
at the origin of the flavour violating processes. 

A first, model-independent approach consists in parametrising cLFV
interactions by means of higher-order non-renormalisable (dimension $d>4$)
operators. The new low-energy effective Lagrangian can be
written as 
$\mathcal{L}^\text{eff} = \mathcal{L}^\text{SM} + \sum_{n \geq 1}
\mathcal{C}_{ij}^{4+n} \, \Lambda^{-n} \, \mathcal{O}_{ij}^{4+n}$, in
which $\Lambda$ denotes the scale of new physics, and $\mathcal{C}$, 
$\mathcal{O}$ the effective couplings and operators, with the former
corresponding to complex matrices in flavour space. Contrary to the
unique dimension-five Weinberg operator (common to all models with
Majorana neutrino masses), there exists a large number of dimension-six
operators, whose low-energy effects include cLFV. 
Regarding the cLFV dimension-six operators, 
these can be loosely classified as dipole, four-fermion and
scalar/vector operators. 

In order to constrain the new physics scale
and the amount of flavour violation thus introduced, the cLFV observables
can be cast (at leading order)
in terms of combinations of $\mathcal{C}_{ij}^{6}$ and 
$\Lambda^{-2}$; simple, natural hypothesis on one allow to infer
constraints on the other. Table~\ref{table:effective} (adapted
from~\cite{Feruglio:2015gka}) collects some bounds on the
scale of new physics (derived under the hypothesis of natural,
$\mathcal{O}(1)$, effective couplings) and on the size of the new
effective couplings (inferred for a choice $\Lambda=1$~TeV). 

{\scriptsize
\renewcommand{\arraystretch}{1.1}
\begin{center}
\begin{table}
\caption{Bounds on the effective couplings and lower bounds on the
  scale $\Lambda$ (TeV), following the hypotheses described on the text; 
the last column refers to the observable leading to the most stringent
bounds.}\label{table:effective}
\vspace*{2mm}\hspace*{3mm}
\begin{tabular}{|c|c|c|c|}
\hline
\begin{tabular}{c}
\, Effective coupling \, \\
(example)
\end{tabular}
& 
\begin{tabular}{c}
\, Bounds on ${\Lambda}$ {(TeV)} \, \\
{(for $|\mathcal{C}^{6}_{ij}| =1$)}
\end{tabular}
& 
\begin{tabular}{c}
\, Bounds on ${|\mathcal{C}^{6}_{ij}|}$ \, \\
{(for $\Lambda = 1$~TeV)}
\end{tabular}
& 
\, Observable \,  \\
\hline
$\mathcal{C}^{\mu e}_{e \gamma}$ & $6.3 \times 10^4$ &
$2.5 \times 10^{-10}$ & $\mu \to e \gamma$ \\
$\mathcal{C}^{\tau e}_{e \gamma}$ & $6.5 \times 10^2$ &
$2.4 \times 10^{-6}$ & $\tau \to e \gamma$ \\
$\mathcal{C}^{\tau \mu}_{e \gamma}$ & $6.1 \times 10^2$ &
$2.7 \times 10^{-6}$ & $\tau \to \mu \gamma$ \\
\hline
$\mathcal{C}^{\mu eee}_{\ell \ell, e e }$ & $207$ & 
$2.3 \times 10^{-5}$ & $\mu \to 3 e $ \\
$\mathcal{C}^{e \tau ee}_{\ell \ell, e e }$ & $10.4$ & 
$9.2 \times 10^{-5}$ & $\tau \to 3 e $ \\
$\mathcal{C}^{\mu \tau \mu \mu}_{\ell \ell, e e }$ & $11.3$ & 
$7.8 \times 10^{-5}$ & $\tau \to 3 \mu $ \\
\hline
$\mathcal{C}^{\mu e}_{(1,3) H\ell}$, $\mathcal{C}^{\mu e}_{He}$ & 
$160$ &
$4 \times 10^{-5}$ & $\mu \to 3 e $ \\
$\mathcal{C}^{\tau e}_{(1,3) H\ell}$, $\mathcal{C}^{\tau e}_{He}$ & 
$\approx 8$ &
$1.5 \times 10^{-2}$ & $\tau \to 3 e $ \\
$\mathcal{C}^{\tau \mu}_{(1,3) H\ell}$, $\mathcal{C}^{\tau \mu}_{He}$ & 
$\approx 9 $ &
$ \approx 10^{-2}$ & $\tau \to 3 \mu $ \\
\hline
\end{tabular}
\end{table}
\end{center}
}
\renewcommand{\arraystretch}{1.}
Despite its appeal for leading to a generic
evaluation of the new physics contributions to a given cLFV observable, 
and thus to model-independent constraints,  
there are several limitations to the effective approach. These include
taking ``natural'' values for the couplings, assuming the dominance of
a single operator when constraining a given process and the
uniqueness of the new physics scale; the latter should be kept in mind when
weighing the impact of the thus derived constraints on new physics. 

\bigskip
A second approach consists in considering
specific new physics models or theories, and evaluating the corresponding
impact for a given class of cLFV processes. 
As extensively explored in the
literature, cLFV might be a powerful test of new physics
realisations, probing scales beyond collider reach, offering valuable
hints on properties and parameters of a given model, and allowing to
disentangle (and ultimately disfavour) between candidate models. 
Interesting examples include generic cLFV extensions of the SM, as is
the case of general supersymmetric (SUSY) models, geometric mechanisms
of cLFV, as in the case of extra-dimensional Randall-Sundrum models,
compositness frameworks (e.g. little(st) Higgs, holographic composite
Higgs, ...), multi-Higgs doublet models, SM extensions via leptoquarks
and/or $Z^\prime$, and finally additional symmetries (be them flavour
or gauge) - of which Left-Right symmetric models and Grand Unified
theories are interesting examples.

A particular appealing class of new physics models regarding cLFV is
that in which all sources lepton flavour violation (neutral and charged) are
related, arising from the mechanism of neutrino mass generation. In what
follows, we will address some examples of the latter, focusing in
realisations of low-scale seesaws and in the supersymmetrisation of a type
I seesaw, further emphasising the r\^ole of cLFV in potentially providing
important information on the mechanism of neutrino mass generation.

\section{cLFV from $\nu$ mass generation}
Although cLFV need not arise from the mechanism of
$\nu$ mass generation, models in which this is indeed the case - such
as the different seesaw realisations - are
particularly appealing and well-motivated frameworks. 
Whether or not a given mechanism of neutrino mass generation does have
an impact regarding cLFV stems from having non-negligble flavour violating
couplings (e.g., the Yukawa couplings) provided that the rates are
not suppressed by excessively heavy propagator masses. While
``standard'' high-scale seesaws do accommodate neutrino data with
natural values of the neutrino Yukawa couplings, the typical scale of the
mediators (close to the GUT scale) leads to a very strong suppression
of the different cLFV rates. On the other hand, 
low-scale seesaws, 
or the embedding of a high-scale seesaw in larger frameworks (as is the
case of the SUSY seesaw), are associated with a rich
phenomenology, with a strong impact regarding cLFV. 

\subsection{Low-scale seesaws}
\begin{figure}   
\begin{center}
\begin{tabular}{cc}
\includegraphics[width=60mm,angle=0]{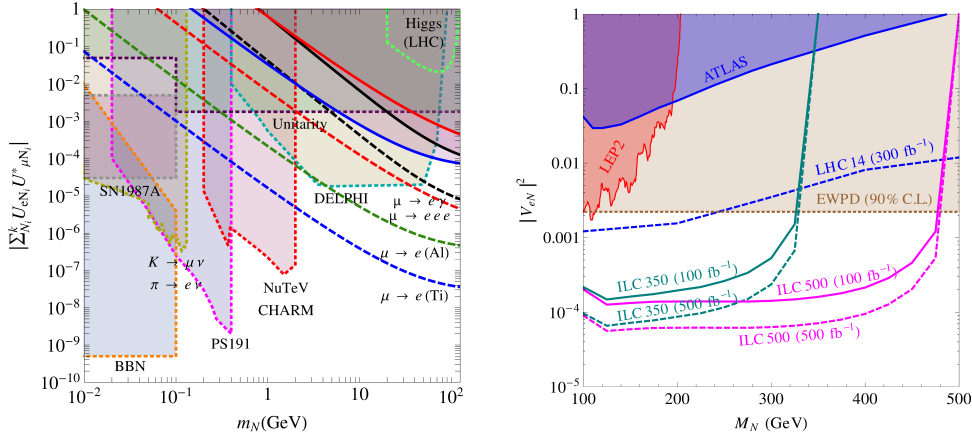} 
\hspace*{0mm}&\hspace*{0mm}
\includegraphics[width=60mm, angle=0]{sensitivity.epsi}
\end{tabular}
\end{center}
\caption[]{
On the left panel, exclusion regions on the heavy state parameter
space (masses and mixings to active states), including current bounds
and future 
sensitivities to $\mu - e $ cLFV observables (from~\cite{Alonso:2012ji}).
On the right, future ILC 
sensitivity reach for the heavy neutrino masses and their mixings, as
well as the currently excluded regions (from~\cite{Banerjee:2015gca}).}
\label{fig:lowscale}
\end{figure}

In low-scale seesaws (as is the case of the
low-scale type I seesaw, inverse and linear seesaw realisations,
...), the new ``heavy'' states do not fully decouple; their non-negligible
mixings with the light (active) neutrinos lead to the non-unitarity of the
left-handed lepton mixing matrix ($U_\text{PMNS} \to \tilde U_\text{PMNS}$),
and thus to having modified neutral and charged lepton currents. The latter
are at the origin of potentially abundant experimental/observational
signatures, which have been intensively searched for in recent years; 
negative results have
allowed to derive strong constraints on the parameter space of the new
degrees of freedom (see~\cite{Alonso:2012ji,Dinh:2012bp} for 
comprehensive discussions of cLFV in low-scale seesaws). An example of
such constraints (presented in the parameter space generated by the
masses of the heavy states and their couplings to the active
neutrinos)~\cite{Alonso:2012ji} is displayed on the left panel of
Fig.~\ref{fig:lowscale}, also including the impact of
$\mu - e $ cLFV observables. Provided that the right-handed neutrinos are  
not excessively heavy, they can be produced in high-energy colliders
and be searched for in several processes (frequently relying on lepton
number violation signatures):
current data (from both LEP and the LHC) already puts strong
constraints on the parameter space; future colliders can further
improve these bounds~\cite{Banerjee:2015gca}, as shown in the right panel of
Fig.~\ref{fig:lowscale}. 

A very appealing example of such low-scale models are Inverse Seesaw (ISS)
realisations: other than right-handed neutrinos, further
sterile states are added; in the case of a (3,3)
ISS realisation, three copies of each are present. 
The masses of the light active neutrinos are
given by a modified seesaw relation, $m_{\nu_i} \approx (Y_\nu v)^2
M_R^{-2} \mu_X$, where $\mu_X$ is the only source of lepton number violation 
in the model. By taking small values of $\mu_X$, one can naturally accommodate
the smallness of active neutrino masses for large Yukawa couplings and
a comparatively low seesaw scale ($M_R$ lying close to the TeV
scale). The spectrum contains, in addition
to the light states, three heavier (mostly sterile) 
pseudo-Dirac pairs, whose masses 
are given by $m_{N} \approx M_R \pm \mu_X$.
 
\begin{figure}   
\begin{center}
\begin{tabular}{cc}
\includegraphics[width=55mm,
  angle=270]{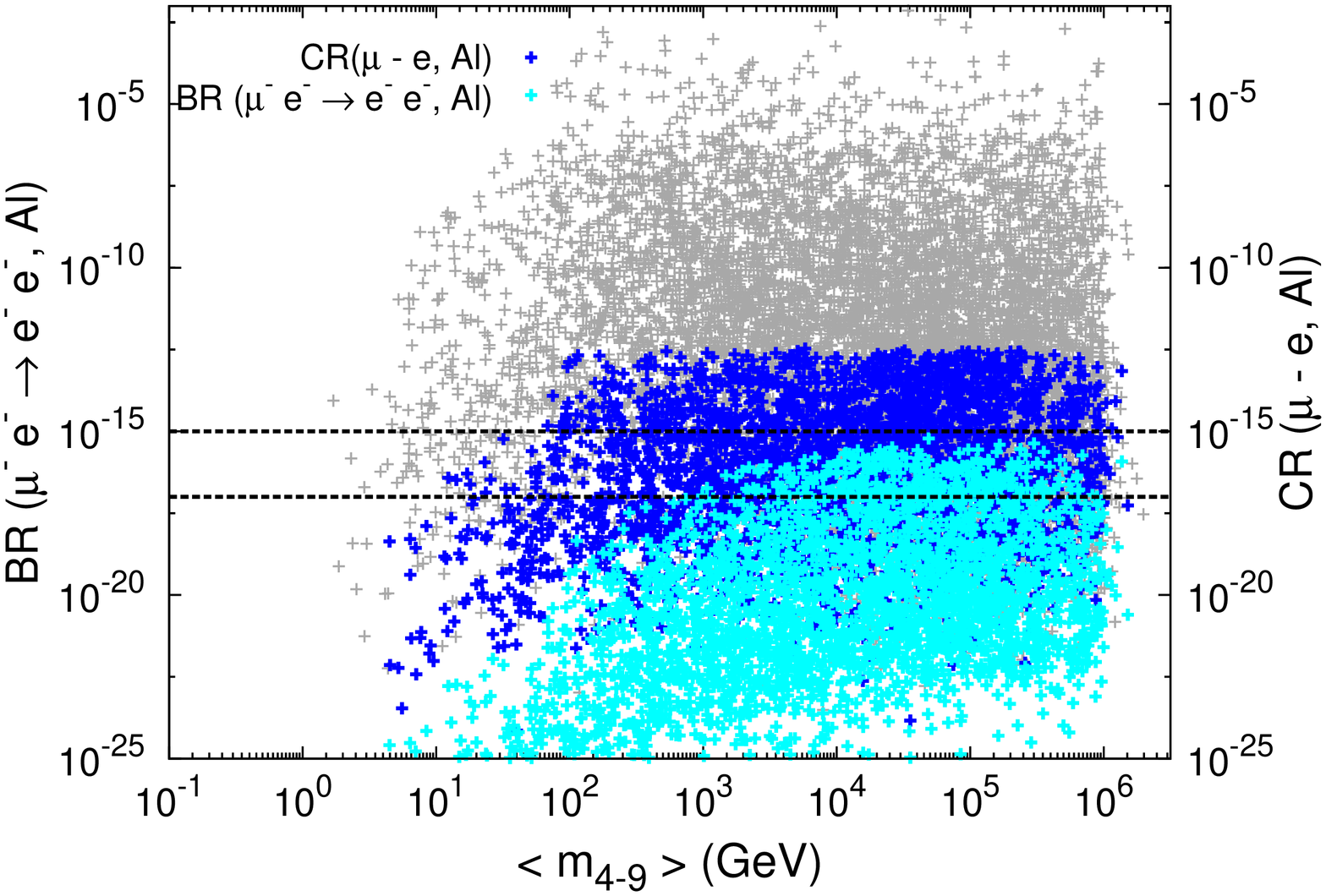} 
\hspace*{0mm}&\hspace*{0mm}
\includegraphics[width=55mm, angle=270]{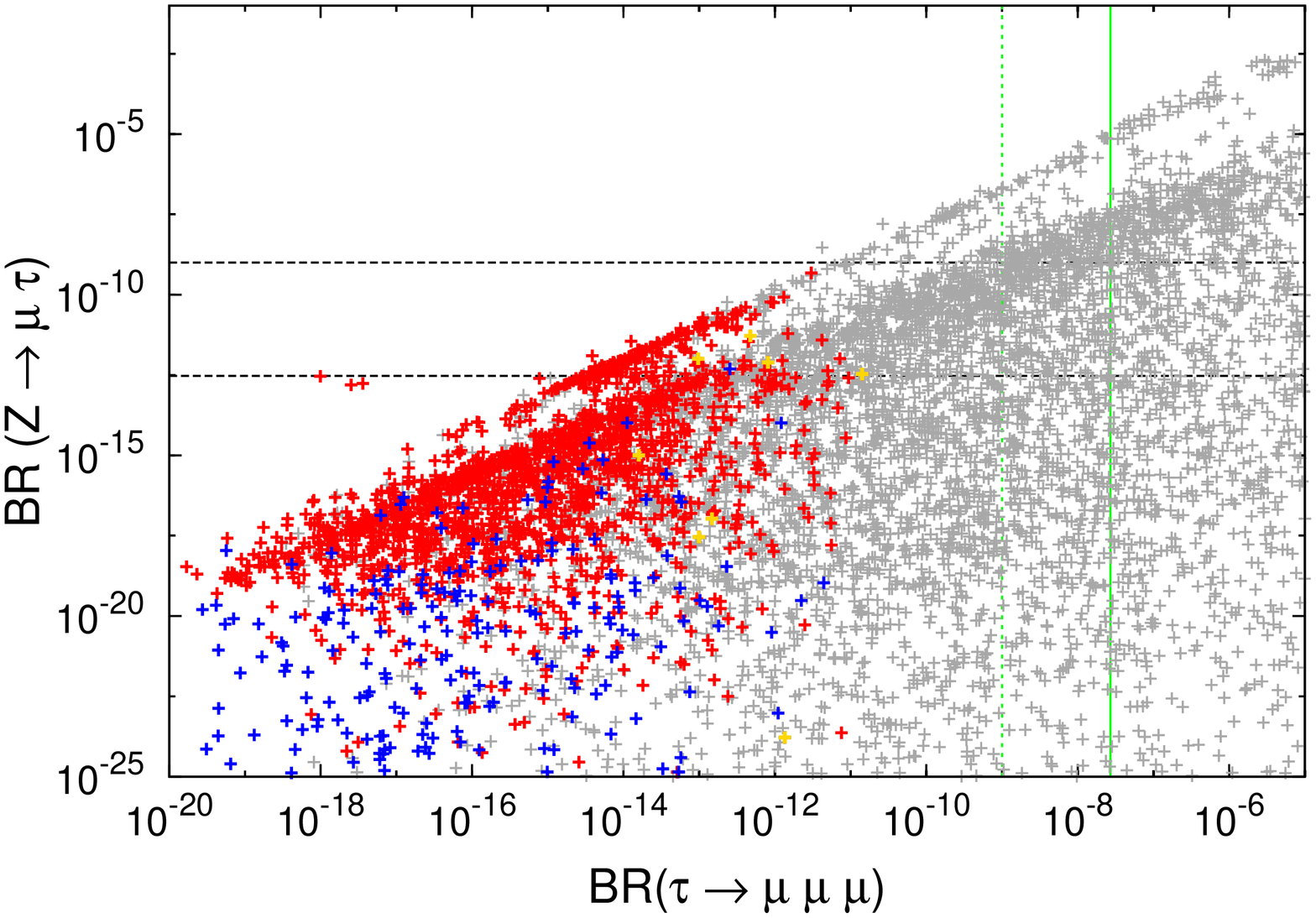}
\end{tabular}
\end{center}
\caption[]{
On the left panel, BR($\mu^- e^- \to e^- e^-$, Al) - cyan -  and CR($\mu-e$, Al)
- blue - as a function of the average mass of the heavier, mostly sterile
states, in a (3,3) ISS realisation. Horizontal lines denote future
experimental sensitivities (from~\cite{Abada:2015oba}).
On the right, BR($Z \to \tau \mu$) vs. BR($\tau \to 3 \mu$) in a (3,3)
ISS realisation. Vertical lines denote future experimental
sensitivities while the horizontal ones correspond to the prospects of
a GigaZ facility and of the FCC-ee (from~\cite{Abada:2014cca}).
In both cases, grey points are phenomenologically excluded.}
\label{fig:iss:low}
\end{figure}

The (3,3) ISS opens the door to a very rich phenomenology, which
includes abundant cLFV signatures, both at low- and at 
high-energies (see, for
example,~\cite{Abada:2013aba,Abada:2015oba,Abada:2014cca}).  
To illustrate the potential impact regarding high-intensity
facilities, the left panel of Fig.~\ref{fig:iss:low} displays the
prospects for $\mu - e$ conversion, as well as the Coulomb enhanced decay of
a muonic atom (both for the case of Aluminium targets), as a function
of the average mass of the heavier states, $<m_{4-9}>$. Although
CR($\mu - e$, Al) is in general associated to
larger rates, for sterile
states above the TeV, both observables are expected to be well within
reach of the COMET experience (horizontal lines respectively denoting
the sensitivity of Phase I and II), or of the Mu2e experiment. 

At higher energies (for example, in the case of a future circular
collider, as FCC-ee), one can also explore cLFV in the decay of
heavier states, as for instance in $Z \to \ell_i \ell_j$. 
In the ISS (3,3) realisation, especially in
the ``large'' sterile mass regime, the cLFV $Z$ decays exhibit a strong
correlation with cLFV 3-body decays (since the latter 
are dominated by the $Z$-penguin
contribution). The prospects for a (3,3) ISS realisation, for the
case of $\mu -\tau$ flavour violation, are
shown in the right panel of Fig.~\ref{fig:iss:low}. Not only
can one expect to have BR($Z \to \tau \mu$) within FCC-ee reach,
but this observable does allow to probe $\mu -\tau$ flavour violation
well beyond the sensitivity of a future SuperB factory (large values
of BR($\tau \to 3 \mu$) are precluded in this realisation due to the
violation of other cLFV bounds).  

At the LHC, searches for heavy ISS mediators relying on cLFV
signatures can be carried; as recently proposed, a significant number
of events (after cuts) could be expected from the channel 
$q q^\prime \to \tau \mu +2$jets (no missing $E^T$)~\cite{Arganda:2015ija}.

\subsection{SUSY type I seesaw}
Another rich and well-motivated framework leading to observable cLFV
is that of the SUSY seesaw (a high-scale seesaw embedded in the
context of otherwise flavour conserving SUSY models). 
In the case of a type I SUSY seesaw~\cite{Borzumati:1986qx}, sizeable 
neutrino Yukawa
couplings (as characteristic of a high-scale seesaw) and the
possibility of new, not excessively heavy mediators (the SUSY partners),
open the door to large contributions to cLFV observables. Having a
unique source of flavour violation implies that the observables
exhibit a high degree of correlation; such a synergy can be explored,
allowing to put the seesaw hypothesis to the test and possibly hinting on
certain parameters. 
For example, the complementarity of two low-energy observables
as is the case of $\mu \to e \gamma$ and $\tau \to \mu \gamma$ has
been explored for different
seesaw scales~\cite{Antusch:2006vw}:
the determination of these two observables, in association with
the discovery of SUSY, would allow to infer
information on the seesaw scale $M_R$, 
or then readily disfavour the SUSY seesaw as the only
source of cLFV. 

High-energy colliders offer direct access to superpartners; the
production of on-shell sleptons would allow to study cLFV in SUSY neutral 
current interactions. There are many cLFV observables that can be
studied - both at the LHC and at a future linear collider. These
include flavoured slepton mass differences 
(the splittings between the first and second
generation charged slepton masses, $\Delta m_{\tilde \ell}$), 
new edges in dilepton mass distributions $m_{\ell \ell}$, and direct
flavour violating final states (in association with decays 
$\chi^0_2 \to \ell_i \ell_j 
\chi^0_1$)~\cite{Abada:2010kj}. A future linear collider would
allow to address further observables - especially should it have the
possibility to operate in $e^- e^-$ mode. In the latter case, the
process $e^- e^- \to \mu^- \mu^- + 2 \chi^0_1$ could become a true
``golden channel'' for cLFV in the present framework~\cite{Abada:2012re}.   

\begin{figure}   
\begin{center}
\hspace*{-5mm}
\begin{tabular}{cc}
\includegraphics[width=67mm]{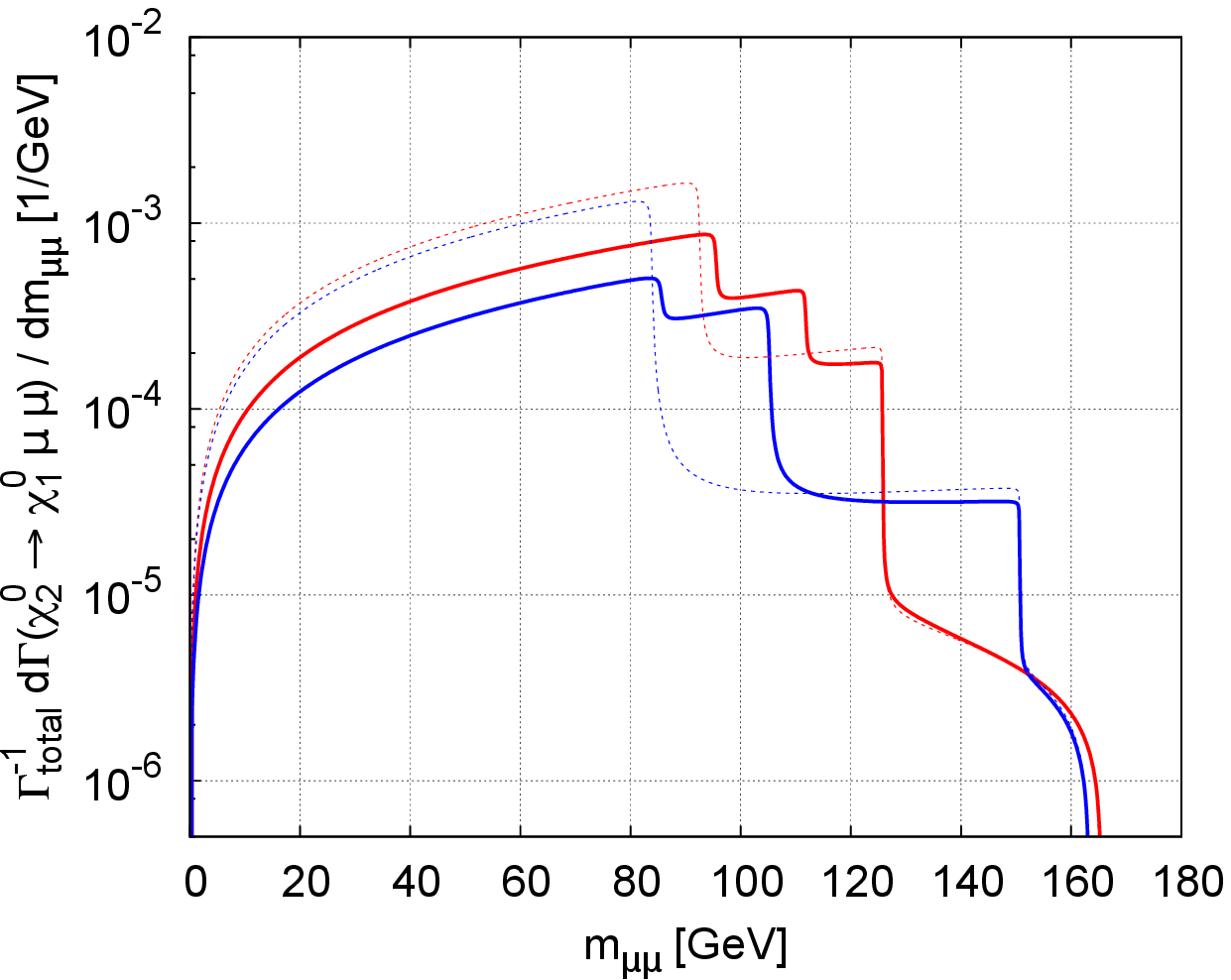}
\hspace*{-2mm}&\hspace*{-3mm}
\raisebox{-4mm}{\includegraphics[width=83mm]{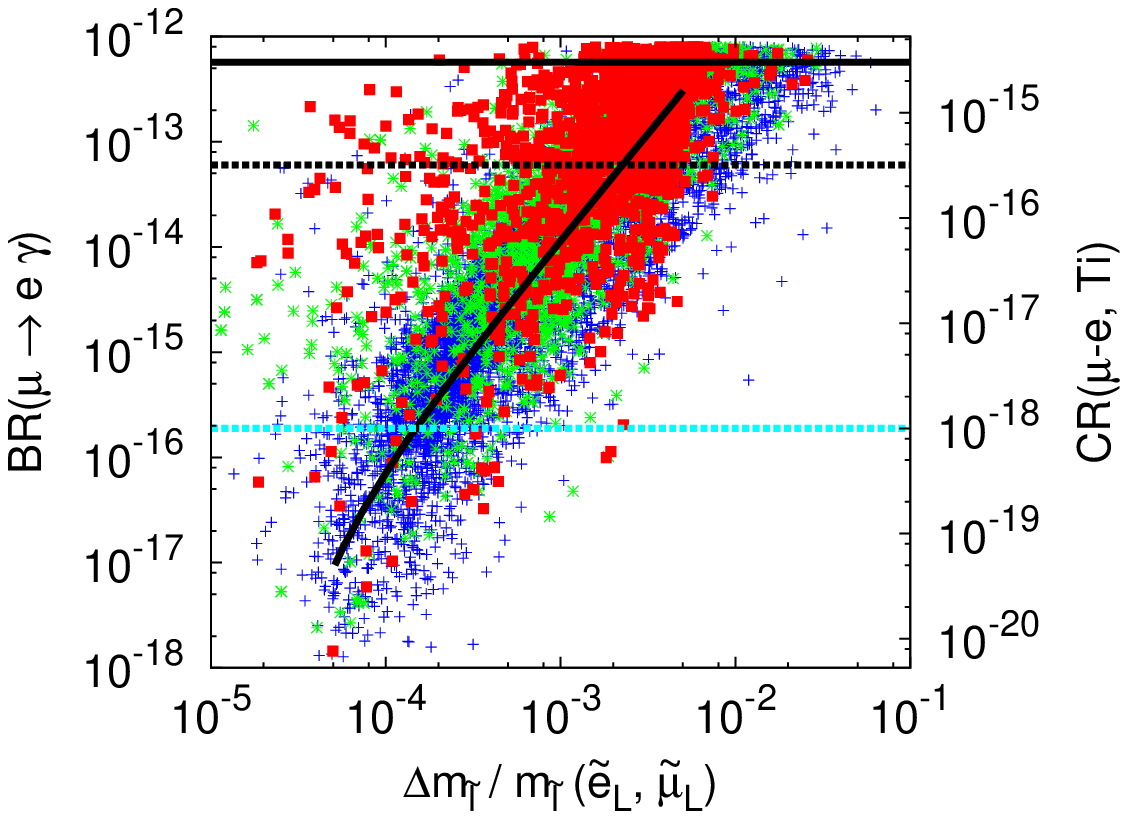}}
\end{tabular}
\end{center}
\caption[]{
On the left, BR($\chi^0_2 \to \mu \mu \chi^0_1$) as a function of 
the dimuon invariant mass $m_{\mu\mu}$ (in GeV) for different SUSY seesaw points.
(from~\cite{Abada:2010kj});
on the right panel, $1^\text{st}$ and $2^\text{nd}$ generation 
charged slepton mass splittings vs. BR($\mu \to e \gamma$), with CR($\mu -e$,
Ti) on secondary y-axis in a type I SUSY seesaw, for different values
of the heaviest 
right-handed neutrino mass $M_{R_3} = 10^{13,14,15}$~GeV ($M_{R_1,R_2} =
10^{10,11}$~GeV) and for a flavour conserving modified mSUGRA benchmark 
(from~\cite{Figueiredo:2013tea}).
\label{fig:susyseesaw}}
\end{figure}

A particularly interesting example is that of new edges in dimuon mass
distributions, which can be studied in association with neutralino
decay cascades at the LHC. In a strictly flavour conserving 
framework, as would be
the case of the constrained Minimal Supersymmetric SM (cMSSM) one is
led to double triangular distributions, with two well-defined edges,
associated with the presence of an intermediate $\tilde \mu_{_L}$ or   
$\tilde \mu_{_R}$ in the decay cascade $\chi^0_2 \to \tilde \mu_{_{L,R}}\, \mu
\to \mu \mu \chi^0_1$ (corresponding to the dashed lines on the left
panel of Fig.~\ref{fig:susyseesaw}). In the flavour violating case of
a type I SUSY seesaw, and in association with other cLFV
manifestations at high and low energies, one observes the appearance
of a third edge (solid lines on the left
panel of Fig.~\ref{fig:susyseesaw}), which reflects that a new state
- a stau - has mediated the neutralino decay:  
$\chi^0_2 \to \tilde \tau_{2} \,\mu
\to \mu \mu \chi^0_1$, thus clearly signalling charged lepton flavour
violation~\cite{Abada:2010kj}. 

The potential of exploring the interplay of
high-intensity (for instance $\mu \to e \gamma$ and  
$\mu-e$ conversion) and other collider
observables (for example, the splittings between left-handed 
selectron and smuon masses, $\Delta m_{\tilde \ell}$) 
is summarised on the right
panel of Fig.~\ref{fig:susyseesaw}: ``isolated'' cLFV manifestations
(i.e., outside the coloured regions) would allow to disfavour the SUSY
seesaw hypothesis as the (unique) underlying source of lepton flavour
violation, while ``compatible'' ones would strenghten it, 
furthermore hinting on the seesaw scale~\cite{Figueiredo:2013tea}. 

\section{Conclusions}
As of today, we have firm evidence that flavour is violated in
the quark sector, as well as in the neutral lepton one. In the absence
of a fundamental principle preventing it, there is no
apparent reason for Nature to conserve charged lepton flavours. By
itself, any observation of a cLFV process would constitute a clear
signal of new physics - beyond the SM extended via massive (Dirac)
neutrinos. 
As we aimed at illustrating in the present brief review, cLFV
observables could provide valuable (indirect) information on the
underlying new physics model, and certainly contribute to at least
disfavour several realisations. 
Interestingly, new physics could even manifest itself indirectly via
cLFV before any direct discovery.

The current (and planned) experimental programme, with numerous
observables being searched for in a large array of high-intensity and
high-energy experiments clearly
renders cLFV a privileged laboratory to search for new physics.

\section*{Acknowledgments}
AMT is gratefull to the Organisers of ``XXVIII Rencontres de Blois''
for the invitation to participate in the Conference.
Part of the work here summarised was done within the framework of the
European Union's Horizon 2020 research and innovation programme under
the Marie Sklodowska-Curie grant agreements No 690575 and No 674896.

\end{document}